\documentclass[twocolumn,groupedaddress]{revtex4}

\usepackage{amsmath,amssymb,amsthm,times,graphics,graphicx,bm}
\usepackage[bookmarks = false, pdfpagemode = None, pdfstartview = FitH, colorlinks = true, urlcolor = blue,hyperfootnotes=false]{hyperref}
\usepackage{color}
\usepackage{dcolumn}

\newcommand{\be}{\begin{equation}}
\newcommand{\ee}{\end{equation}}
\newcommand{\ben}{\begin{eqnarray}}
\newcommand{\een}{\end{eqnarray}}
\def\bra#1{\mathinner{\langle{#1}|}}
\def\ket#1{\mathinner{|{#1}\rangle}}

\def\tr{ {\rm{Tr }}}

\begin{document}

\title{Highly efficient energy excitation transfer in
light-harvesting complexes:\\ The
fundamental role of noise-assisted transport}

\author{F. Caruso$^{1,2}$}
\author{A.W. Chin$^{3}$}
\author{A. Datta$^{1,2}$}
\author{S.F. Huelga$^{3}$}
\author{M.B. Plenio$^{1,2}$}

\affiliation{$^{1}$Institute for Mathematical Sciences, 53
Prince's Gate, Imperial College, London, SW7 2PG, UK}
\affiliation{$^{2}$QOLS, The Blackett Laboratory, Imperial
College, London, Prince Consort Road, SW7 2BW, UK}
\affiliation{$^{3}$Quantum Physics Group, Department of Physics,
Astronomy \& Mathematics, University of Hertfordshire, Hatfield,
Herts AL10 9AB, UK}

\begin{abstract}
Excitation transfer through interacting systems plays an
important role in many areas of physics, chemistry, and biology.
The uncontrollable interaction of the transmission network with a
noisy environment is usually assumed to deteriorate its transport
capacity, especially so when the system is fundamentally quantum
mechanical. Here we identify key mechanisms through which noise
such as dephasing, perhaps counter intuitively, may actually aid
transport through a dissipative network by opening up additional
pathways for excitation transfer. We show that these are processes
that lead to the inhibition of destructive interference and
exploitation of line broadening effects. We illustrate how these
mechanisms operate on a fully connected network by developing a
powerful analytical technique that identifies the invariant (excitation trapping) subspaces of a given Hamiltonian. Finally, we show how these principles can explain the remarkable efficiency
and robustness of excitation energy transfer from the light-harvesting
chlorosomes to the bacterial reaction center in photosynthetic
complexes and present a numerical analysis of excitation transport
across the Fenna-Matthew-Olson (FMO) complex together with a brief
analysis of its entanglement properties. Our results show that, in
general, it is the careful interplay of quantum mechanical features
and the unavoidable environmental noise that will lead to an optimal
system performance.

\end{abstract}

\maketitle

Transport phenomena have been central to quantum mechanics
since its early days \cite{perrin32}. This pivotal role has been
recently renewed by the prospect of transferring quantum
information across quantum networks \cite{plenio04} and the
recurring interest in understanding the fundamental processes
influencing energy transport in photosynthetic molecules
~\cite{sension07}. Light harvesting complexes can harness the
available light energy at efficiencies well above $90\%$ in
conditions that are often defined as {\em hot and wet}
\cite{nagy,ross}.
Understanding the workings of such a process has the potential to
be immensely valuable from a technological point of view, potentially paving the way for the design of novel nanofabricated
structures for quantum transport and optimized solar cells.
There is already an important body of work analyzing energy transport in these systems
\cite{forster48,redfield65,groversilbey71,kenkre74,scholes03,adolphs06}. However, the clear identification of
the basic mechanisms underlying this remarkably efficient natural
transfer scheme has remained elusive. Recently, a sequence of beautiful experiments using very
effective nonlinear spectroscopic techniques
probed the dynamics of delocalized exciton states in
light-harvesting complexes \cite{fleming07b}, the
Fenna-Matthew-Olson (FMO) complex~\cite{phna02,fleming07a} and
conjugated polymer samples \cite{scholes}, the latter modeling
multichromophoric systems. Based on these experimental
observations, quantum coherence across multiple chromophoric
sites has been suggested as the probable cause of the highly
efficient energy transfer in photosynthetic systems. The influence of initial exciton delocalization in the
efficiency of excitation transfer was investigated by \cite{olaya-castro08} and it has been suggested
that the enhanced transfer rates may be
attributed to the exploitation of principles of quantum search
algorithms by the quantum dynamics of the FMO complex
\cite{sension07,fleming07a}. At the same time, however,
excitations in chromophoric complexes are subject to strong
dephasing noise from a quasi-continuum of environmental phonon
modes, which, in turn, suggests that quantum coherence will be
short-ranged rather than extending across the entire complex. This
short ranged nature of quantum correlations may suggest that
concepts from quantum computation will not play as decisive
a role as it is generally accepted. In fact, efficient pure
state quantum computation requires long range entanglement
\cite{jozsa}.  It therefore remains unclear whether the
presence of quantum entanglement has any specific functional
role in the process of energy transport across biological
systems, or it can simply be viewed as the unavoidable
by-product of the existence of a coherent quantum evolution
during a certain time scale in such systems.

         Inspired by these experimental results and their
interpretations, further theoretical studies of energy transport in light
harvesting complexes have been carried out recently
\cite{aspuruguzik08,plenio08}. They investigate the role of noise,
and in particular dephasing, in the process of exciton transport in
these complexes. Contrary to the conventional wisdom that noise
always deteriorates the performance of a system, several instances
are known, in classical as well as quantum mechanics, where noise
is known to be of advantage in enhancing the system's response. In
the classical phenomenon of stochastic resonance
\cite{gammaitoni98}, a non-linear system may benefit from the
presence of noise to achieve an enhanced sensitivity to a weak
signal. Motivated by this observation, it was successfully
demonstrated that noise may have the capacity to generate
genuinely quantum mechanical features such as quantum coherence
and entanglement \cite{plenio02,plenio07,others}. Taking this further, it was
shown recently that noise, in the form of local dephasing, has the
ability to enhance the rate and efficiency of energy transfer when
compared to a perfectly quantum coherent
system~\cite{aspuruguzik08,plenio08}. Indeed, the highly effective
exciton transfer in light harvesting complexes requires the
simultaneous presence of quantum coherent evolution and dephasing
noise. Simulations have also shown that, in the competition
between coherent dynamics and incoherent energy dissipation
(population decay), the former regime lasts for about
several hundreds of fs ~\cite{Mukamel08} at the beginning of the
exciton transfer process that spans about
$5\;\mathrm{ps}$~\cite{adolphs06}. As a result, in addition to the
expectation that the quantum coherence is short-ranged, it is in
fact short-lived as well, and so will be any entanglement across the systems.

In spite of these qualitative and quantitative successes based
on numerical work, the underlying mechanisms by which noise
supports transport have yet to be precisely identified and
exemplified in simple models. Achieving this may assist us
in replicating in artificial structures what nature does so
well - energy transfer in light-harvesting complexes.

In this paper we identify two fundamental mechanisms underlying
dephasing-assisted transport and elucidate them in the context
of two basic models. We will show that these are firstly, the
suppression of destructive interference by adding local decohering
noise such as dephasing or static disorder, and secondly, the
enhancement of transfer by line broadening caused, for example,
by the fluctuation of energy levels. Both mechanisms
have the principal effect of opening additional channels for
transport in the system; channels that would be inhibited
under solely coherent evolution.

We start by presenting in section I an abstract network model
for quantum transport where local sites are subject to both local
dissipation and dephasing and the excitation transfer to a
reaction center is modeled via an irreversible coupling to a
privileged trapping site, an excitation {\em sink}. In section
II we introduce a simplified model in terms of a fully connected
network, where all sites are coupled to each other with equal
coupling strength. This model will allow us to develop a powerful
analytical technique to elucidate whether a given Hamiltonian is
susceptible to support noise assisted transport. This will be
done by characterizing the invariant (excitation trapping)
subspaces of the Hamiltonian. With the aim of making the pace of the paper reasonably
fluid, the detailed procedure is
presented in a separate Appendix where we develop a technique
for solving the complete master equation for the fully connected
network. Using these analytical results, we show in sections
II.A-II.D that dephasing assists transport in fully connected
networks while dissipation only does so in some cases.
We provide examples when pure dissipation may assist the
transport through a network. Section II.E revises
line broadening as a classical effect. In section III we extend
the model to account for forms of spatially correlated noise.
This theoretical analysis allows for the identification of the
fundamental mechanisms
underlying dephasing-assisted transport which are then applied in section IV to the specific example of energy transfer across a light harvesting system, the FMO complex, which may
be modelled as a network of seven nodes, fully connected, albeit
with non-uniform coupling strengths. We show how experimental results concerning exciton transfer cannot be reproduced in the absence of dephasing while the inclusion of this form of local noise boosts energy transfer towards the observed values in the correct time scale.
We show that the two mechanisms
presented here by which dephasing assists excitation transfer
are very robust against variations of the detailed structure
of the system.
The theoretically expected and numerically observed weak
dependence on the noise strength suggests that experimental
results which have been obtained at 77 K should remain broadly
valid at higher temperature~\cite{fleming07a} and that these
processes will be robust against perturbations due to different
environments.
We conclude section IV with an evaluation of
entanglement generation and transmission within this simple noise model, including local and correlated dephasing, and show that quantum correlations are expected to be confined to the initial steps of the transport process and die out well before the excitation is fully transferred.
In section V, we summarize the main physical ideas behind our
approach in an intuitive form and we conclude in section VI by suggesting that the principles outlined
here may find broad applicability in understanding the dynamics of
biological complexes and transport phenomena and discuss future
work in this direction.

\section{The Network Model}
Light-harvesting complexes are typically constituted of multiple
chromophores which transform photons into excitons and transport
them to a reaction center \cite{cheng00}. Experimental studies of
the exciton dynamics in such systems reveal rich transport
dynamics consisting of short-time coherent quantum dynamics which
evolve, in the presence of noise into an incoherent population
transport which irreversibly transfers excitations to the reaction
center. In order to elucidate the basic phenomena clearly without
overburdening the description with detail, we consider the relevant
complexes as systems composed of several distinct sites, one of
which is connected to the chromosomes while another is connected
to the reaction center.
This complex effective dynamics will then
be modelled by a combination of simple Hamiltonian dynamics which
describe the coherent exchange of excitations between sites, and
local Lindblad terms that take into account the dephasing and
dissipation caused by the external environment.
A network of $N$ sites will be described by the Hamiltonian,
\begin{equation}\label{Hamiltonian}
        H = \sum_{j=1}^N \hbar\omega_j \sigma_j^{+}\sigma_j^{-}
        + \sum_{j\neq l} \hbar v_{j,l} (\sigma_j^{-}
        \sigma_{l}^{+} + \sigma_j^{+}\sigma_{l}^{-}),
\end{equation}
where $\sigma_j^{+}$ ($\sigma_j^{-}$) are the raising and lowering
operators for site $j$, $\hbar\omega_j$ is the local site
excitation energy and $v_{k,l}$ denotes the hopping rate of an
excitation between the sites $k$ and $l$. We will also designate a
site $0$ representing the zero exciton state of the complex, which
appears in operators such as $\sigma_j^+ = \ket{j}\bra{0}$, where
the state $|j\rangle$ denotes one excitation in the site $j$. We
will assume that the system is susceptible simultaneously to two
distinct types of noise processes, a dissipative process that
transfers the excitation energy in site $j$ to the environment
(with rate $\Gamma_j$) and a pure dephasing process (with rate
$\gamma_j$) that destroys the phase coherence of any superposition
state in the system, i.e randomizing the local excitation phase in
site $j$. Following previous studies on dephasing-assisted
transport \cite{plenio08,aspuruguzik08}, we describe both
processes using a Markovian master equation with local dephasing
and dissipation terms \cite{note1}. As we shall see, the
mechanisms that we discuss here are quite insensitive to the
details of the noise model. In the Markovian master equation
approach, the dissipative and the energy-conserving dephasing
processes are captured, respectively, by the Lindblad
super-operators
\begin{eqnarray}
        \label{dissipation}
        {\cal L}_{diss}(\rho) &=& \sum_{j=1}^{N} \Gamma_j[
        -\{\sigma_j^{+}\sigma_j^{-},\rho\} +
        2 \sigma_j^{-}\rho \sigma_j^{+} ], \\
        {\cal L}_{deph}(\rho) &=& \sum_{j=1}^{N} \gamma_j[
        -\{\sigma_j^{+}\sigma_j^{-},\rho\} +
        2 \sigma_j^{+}\sigma_j^{-}\rho \sigma_j^{+}\sigma_j^{-}]\;.
\end{eqnarray}
Finally, the total transfer of excitation is measured by the
population in the `sink', numbered $N+1$, which is populated by an
irreversible decay process (with rate $\Gamma_{N+1}$) from a
chosen site $k$ as described by the Lindblad operator
\ben
        {\cal L}_{sink}(\rho) &=& \Gamma_{N+1}[2\sigma_{N+1}^{+}\sigma_k^{-}
        \rho \sigma_k^{+}\sigma_{N+1}^{-} - \nonumber \\
        && \hskip1.0cm\{\sigma_k^{+}\sigma_{N+1}^{-}\sigma_{N+1}^{+} \sigma_k^{-},\rho\} ].
\een
For definitiveness and simplicity, the initial state of the
network at $t=0$ will be assumed to be a single excitation in site
$1$ (i.e., state $|1\rangle$), unless stated otherwise. The model
is completed by introducing the quantity by which we measure the
efficiency of network's transport properties, that is the
population transferred to the sink $p_{sink} (t)$, which is given
by $p_{sink}(t) = 2\Gamma_{N+1}\int_{0}^t\rho_{kk}(t')\mathrm{d}t'$.
\begin{figure}[t]
\centerline{\includegraphics[width=.45\textwidth]{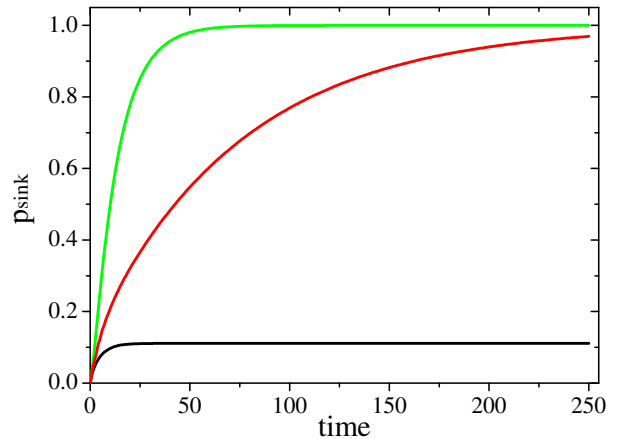}}
\caption{$p_{sink}$ vs. time is shown for a fully connected graph
of $N=10$ nodes with $\Gamma_i=0$ for $i=1,\ldots N$, $J=1,
\Gamma_{N+1}=1$. At $t=0$ one excitation is in the site $1$ and
the energy transfer evolves according to the quantum dynamics
investigated in the text. The cases of an energy mismatch, i.e.
$\omega_1=1, \omega_{i\neq 1}=0$ (red line), a dephasing mismatch,
i.e. $\gamma_1=1,$ $\gamma_{i\neq 1}=0$ (green line), and the
basic case $\gamma_i=\omega_i=0$ (black line) $\forall i$ are
shown. For the latter, the population in the sink asymptotically
reaches $p_{sink}=1/9$. Note that both the red and the green lines
reach unit transfer asymptotically, but at considerably different
rates.}\label{fig3}
\end{figure}
\section{The Fully Connected Network (FCN)}
The FMO complex can be described by a Hamiltonian in which every
site in the network is coupled to every other sites but with site
dependent coupling strength \cite{plenio08,adolphs06}. The
dominant couplings are the nearest neighbour terms, but
significant hopping matrix elements exist also between more
distant sites. This motivates the study of networks with a high
level of connectivity (see also Ref. \cite{tsomkos} for quantum
state transfer), and in this section, for simplicity, we will look
in detail at a fully connected network (FCN). The FCN is
characterised by equal hopping strengths between all sites, i.e.
$\hbar v_{j,l}=J$ for any $j \neq l$. Remarkably, for the case of
a uniform network, i.e. one in which $\omega_{j}$, $\gamma_{j}$,
and $\Gamma_{j}$ are the same on every site, an exact analytical
solution can be found for the density matrix of arbitrarily large
networks. These exact solutions are obtained by defining a set of
collective variables, and the formal development of the solutions
is given in the Appendix. These exact solutions provide insight
into the real-time transport dynamics of the FCN, and, by
considering different scenarios in which various noise effects are
present or absent, we can isolate the various mechanisms that
contribute to the dephasing-assisted transport, as discussed
below. Moreover, having isolated these mechanisms, we find that
they can be described in an intuitive wave mechanics picture which
highlights the relevance of these processes in a much wider range
of quantum network models.
\subsection{No dissipation, no dephasing: Destructive interference}
\begin{figure}[b]
\centerline{\includegraphics[width=.26\textwidth]{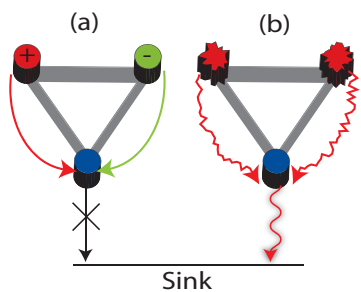}}
\caption{A fully-connected three-site network. In (a) the
excitation wave function is delocalised over two sites (red and
green) with equal probability of being found at either site.
However, as the wavefunction is antisymmetric with respect to the
interchange of red and green, this state has no overlap with the
dissipative site (blue) due to the destructive interference of the
tunnelling amplitudes from each site in the superposition. The
network can therefore store an excitation in this state
indefinitely. In (b) pure dephasing causes the loss of this phase
coherence and the two amplitudes no longer cancel, leading to
total transfer of the excitation to the sink.}\label{fig4}
\end{figure}

This case is described by $\gamma_j = \Gamma_j = 0$ and, for
simplicity, $\omega_{j} = 0$ for any $j=1,\ldots,N$. The only
irreversible process left in this system is the decay of
population to the sink from site $N.$ The exact solution (in the
Appendix) predicts the striking result that as $t\rightarrow
\infty$ the total amount of the excitation that is transferred to
the sink is given by
 \be
p_{sink}(\infty)=\frac{1}{N-1}\;.
 \ee
In Fig.~(\ref{fig3}), the time evolution of the sink population is
shown (black line), for the case of $N=10$. For any reasonably
large network the transfer to the sink is very small, even in the
limit $t\rightarrow\infty$. This should be contrasted with
classical hopping, e.g. a random walk, model, in which the
excitation can be shown to be completely transferred to the sink
as $t\rightarrow\infty$. The difference between these results is a
consequence of the wave-like nature of the quantum dynamics in the
FCN and the presence of destructive interference. This can be seen
by considering the final (i.e., $t\rightarrow\infty$) state of the
network given by
\begin{equation}
        \label{state}
        \rho(t\rightarrow\infty) =
        \frac{N-2}{N-1}|\Psi\rangle\langle\Psi| + \frac{1}{N-1}|N+1\rangle\langle N+1|
\end{equation}
where
\be
|\Psi\rangle=\frac{1}{\sqrt{(N-1)(N-2)}}\sum_{j\neq N,1}(|1\rangle-|j\rangle),
 \ee
and $|1\rangle$ is the initially occupied site. Even though each
individual site has a finite amplitude $J$ for transfer to site
$N$, the amplitudes coming from $|1\rangle$ cancel those from
$|j\rangle$ due to the minus sign in the superposition. As a
result of this {\em destructive interference}, the net amplitude
connecting $|\Psi\rangle$ and $|N\rangle$ must be zero, which
means that the excitation stored in $|\Psi\rangle$ is unaffected
by any process that acts locally at $N$. Thus a quantum FCN can
protectively store some of the excitation in superpositions that
have no effective overlap with site $N$, as illustrated in Fig.
(\ref{fig4}(a)).
A powerful way of understanding how this final state emerges
without actually solving the full dynamics is to invoke the notion
of an invariant subspace. We define an invariant subspace to
consist of a set of states which are both eigenstates of $H$ and
have no overlap with site $N$. The eigenstates of the FCN consist
of a state of the form
$|\phi\rangle=N^{-\frac{1}{2}}\sum_{j=1}^{N}|j\rangle$ with energy
$E=J(N-1)$ and $N-1$ degenerate eigenstates with $E=-J$. Due to
the degeneracy of these states, they can be represented in several
ways. One particularly useful representation consists of
anti-symmetric superpositions of pairs of sites of the form
$|\psi_{j}\rangle=|1\rangle-|j\rangle$ where $j=2 \dots N$. We now
expand the initial system state in terms of the invariant subspace
and the remaining states to find
 \be
 \label{expansion}
\ket{1}=\left(\frac{1}{N-1}\right)\left[\overbrace{\left(
\sum_{j=2}^{N-1}(|\psi_{j}\rangle\right)}^{\mathrm{Invariant}}+
\sqrt{N}|\phi\rangle
-|N\rangle\right].
 \ee
The subsequent time evolution of the system is very simple. The
states in the invariant subspace are eigenstates of the Hamiltonian
and are not affected by the open-system dynamics that only acts at
site $N$. Therefore, their evolution is purely coherent and, being
degenerate, can be described by a simple global phase. The remaining
components of the initial state expansion evolve in the
non-invariant subspace defined by having a finite overlap with site
$N$ and are therefore transferred into the sink. The net result is
that at long times the weight held in the network is simply that
contained in the invariant part of the initial state expansion. A
further consequence of this analysis is the prediction that, if the
system is initially prepared in one of the invariant states, then
there are no dynamics: the excitation is completely trapped in a
noise-protected stationary state and $p_{sink}(\infty)=0$, as
illustrated in Fig.~(\ref{fig4}(a)). Note that this phenomenon is
reminiscent of coherent population trapping in quantum optical
\cite{trap1} and condensed matter \cite{trap2,trap3} systems.

The power of this approach is that the argument given above must
also be valid for other Hamiltonians provided they contain an
invariant subspace. Invariant subspaces turn out to be rather easy
to find, and can be systematically generated for any network with
some degeneracy. This would naturally imply that large invariant
subspaces can be found in systems of high symmetry, and the larger
the invariant subspace is, the larger is the spectral weight
retained by the network. A brief discussion of more general
networks within the invariant subspace approach can be found in
the Appendix. These results are independent of the noise model
provided that it only acts locally at $N$.  However, under
non-local noise it may also be possible to find invariant
subspaces as well.
\subsection{Static disorder: Suppressing destructive
interference}
Another advantage of the invariant subspace approach is that it
can be extended to situations in which the uniformity of the
network is perturbed and allows us to study one of the key
mechanisms for suppressing destructive interference. Changes to
the local site energies are local perturbations, and it should
once again be possible to construct an invariant subspace that
does not feel their influence. This requires that we define a new
invariant subspace that is spanned by all eigenstates that do not
have any overlap with either site $N$, connected to the sink, or
the perturbed sites. For the FCN this amounts to a reduction of
the previous invariant subspace by the number of perturbed sites,
that is, we now exclude all $|\psi_{j}\rangle$ corresponding to
the perturbed sites $j$. The final state of the system now
consists of the part of the initial state expansion that lies in
the reduced invariant subspace, which for FCN immediately implies
\begin{equation}
        p_{sink}(\infty) = \frac{1}{N-D-1}
\end{equation}
where $D$ is the number of sites with different site energies
\cite{note2}. This implies that a network where all the site
energies differ ($D=N-2$) results in $p_{sink}(\infty)=1$, i.e.
complete excitation transfer. A natural consequence of this result
is that a network with randomly disordered energies always leads
to perfect transfer as $t\rightarrow \infty$.

While in the FMO complex and biological complexes in general other
mechanisms may also play a role, the above strongly suggests that it is
not an accident that in many of those complexes a moderate amount of
disorder is present in the distribution of the site energies.

\begin{figure}[b]
\includegraphics[width=.45\textwidth]{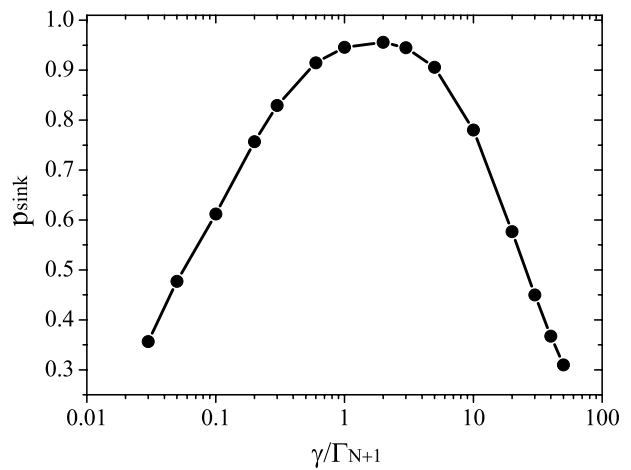}
\caption{Dependence of $p_{sink}(t)$ at a fixed time $t=20$ as a
function of $\gamma/\Gamma_{N+1}$, in the case of $N=5$, $J=1$,
and $\omega=0$. The initially sharp rise is due to the increasing
rapidity at which the invariant subspace is destroyed, whilst the
decreasing part is due to quantum Zeno effects.} \label{fig5a}
\end{figure}

The significance of this observation should not be underestimated.
It contrasts with the idea that disorder leads to either weak or
Anderson localization in the system \cite{Anderson58,lee85}, and
that static disorder always inhibits transfer. That this is not the
case in the FCN is already demonstrated by the above results where
at least a certain amount of disorder is essential for achieving
complete population transfer. The origin of this perhaps surprising
behavior is that, unlike systems described by the Anderson tight-binding
Hamiltonian, the FCN transport to the sink is already strongly
suppressed in
the absence of disorder due to destructive interference of transition
amplitudes to the state coupled to the sink. As we have shown, this
destructive interference is inhibited by any finite amount of disorder
and thus disorder can initially lead to an enhancement of transport
in the FCN. However, while finite disorder leads to $p_{sink}(\infty)=1$,
the rate at which the excitation is transferred to the sink is not
a monotonic function of the disorder strength - see Fig. (\ref{fig5a}).
For weak disorder
the spectrum of the FCN still contains states that are fairly
close to the uncoupled subspace, i.e. they couple only weakly
to the sink, and transport through the network will be slow.
As the size of the disorder increases, the eigenstates become
increasingly distant from the uncoupled subspace and should
lead to an increase in the transport rate. Increasing the magnitude
of the disorder further eventually causes the transfer through
the network to become slower again due to the reduction in overlap
between the sites, as will be discussed later on.
\subsection{Local dephasing without dissipation: Suppressing
destructive interference}
The previous sections showed that the noise-free FCN can trap a
large part of the initial excitation within a manifold of states
that do not overlap with site $N$ and hence do not lead to
transfer to the sink. These states are coherent superpositions of
excitons in different sites, and are invariant in time due to {\em
destructive interference} of the transition amplitudes. Static
disorder however may suppress this destructive interference, thus
releasing all or part of the trapped excitons. Here we show that
local dephasing noise has a very similar effect because it affects
the relative phase in the states that are invariant under the
Hamiltonian dynamics, e.g. mapping the invariant state
$(|1\rangle-|2\rangle)/\sqrt{2}$ to the evolving state
$(|1\rangle+|2\rangle)/\sqrt{2}$
(in the master equation picture presented above for example
through the action of the operator $\sigma_2^+\sigma_2^-$), and
thus perturbing the important destructive interference. This
observation is quite general and may be made for arbitrary
dephasing mechanisms both due to classical or quantum environments.
Indeed, in
the presence of local dephasing on all sites the size of the
invariant subspace discussed there vanishes and one finds that as
$t\rightarrow\infty$, $p_{sink}$ tends to unity for any finite
$\gamma$. This may be confirmed analytically for the case of a
uniform local dephasing rate $\gamma_{j}=\gamma$, for which the dynamical
equations can again be solved exactly, as shown in the Appendix.
\begin{figure}[t]
\includegraphics[width=.45\textwidth]{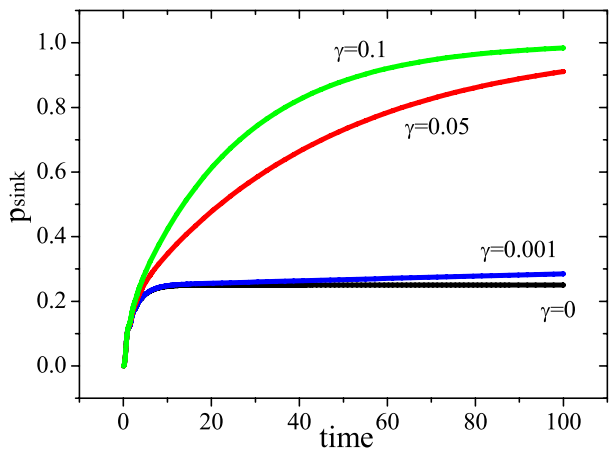}
\caption{Population of the sink $p_{sink}(t)$ as a function of
time $t$ for increasing dephasing rates $\gamma$, in the case of
$\Gamma_{N+1}=1$. For short times, the system dynamics are
identical to the case of zero dephasing, and for very weak
dephasing a significant fraction of the excitation can be trapped
in invariant subspace for a time of approximately $\gamma^{-1}$.}
\label{fig5b}
\end{figure}
This demonstrates the same fundamentally quantum mechanical aspect
of dephasing-assisted transport as the Mach-Zender interferometer in Sec. V.A:
the removal of destructive interference and the creation of new
pathways for transport (see Fig.~(\ref{fig4}(b))). Given that all
physical realisations of network models are subject to some
dephasing, one might wonder why we have laboured to develop the
notion of invariant subspaces, spaces which are anything but
invariant in a realistic network. One reason is that the dynamics
of the excitation transfer can still be strongly influenced by the
presence of destructive interference. This is shown in
Fig.~(\ref{fig5b}) which shows the evolution of $p_{sink}(t)$ for
$N=5$ and $J=1$ as a function of time for various pure dephasing
rates. For $\gamma\ll\Gamma_{N+1}$ the early time evolution is
identical to the case of no dephasing. This implies that the
system evolves into the final state found in the last section and
then slowly dephases leading to a gradual transfer of the
remaining spectral weight to the bath. Interestingly, the speed at
which the excitation is transferred is not a monotonic function of
the dephasing rate. This is illustrated in Fig.~(\ref{fig5a})
which shows $p_{sink}(t)$ as a function of $\gamma/\Gamma_{N+1}$ at a
fixed time of $t=20$. The initial sharp increase in the transport
rate is due to the destruction of the invariant subspace as
discussed above, whilst the slowing of the transfer rate for large
dephasing rates can be ascribed to either a quantum Zeno effect
or, equivalently, an effect of line broadening. We will discuss
the latter in more detail below.

In Fig.~(\ref{fig3}), we compare our base case
($\gamma=\omega=\Gamma=0$) to two other cases both of which lead
to $p_{sink}=1.$ One is the case of just a local dephasing on the
site of injection ($\gamma=0$ elsewhere), while in the other case
we have a different site energy at the site of injection
($\omega=0$ elsewhere). We can see very clearly that pure
dephasing leads to a faster transfer than pure energy mismatch. An
insight of this nature could help us control the transport
behaviour in laboratory and real systems much more effectively. As
shown above by using the invariant subspace argument, one may
increase the transfer to the sink in small controlled steps, thus
carrying us to a limit where we can almost direct the flow of
energy in systems at the microscopic level by engineering
macroscopic parameters.
\subsection{Only local dissipation, no dephasing}
Dissipation of the exciton by the environment also leads to the
suppression of destructive interference, which raises the natural
question whether relaxation alone can enhance transport in the
absence of pure dephasing or static disorder of site energies. The
answer depends on the system and initial state. To address this
question, we consider the case of $\gamma_j=0$ and
$\Gamma_{j}=\Gamma$. The exact solution is presented in the
Appendix and predicts
\begin{equation}
\label{psinkdiss}
        p_{sink}(\infty)=\frac{J^{2}\Gamma_{N+1}(2\Gamma+\Gamma_{N+1})}
        {\phi(\Gamma,\Gamma_{N+1},J,N)},
\end{equation}
where $\phi$ is given by $\phi(\Gamma,\Gamma_{N+1},J,N)=
4\Gamma^{4}+8\Gamma^{3}\Gamma_{N+1}+
5\Gamma^{2}\Gamma_{N+1}^{2}+\Gamma\Gamma_{N+1}^{3}
+\Gamma_{N+1}^{2}J^{2}(N-1)+\Gamma^{2}J^{2}N^{2}+\Gamma\Gamma_{N+1}J^{2}N^{2}$.
From Eq. (\ref{psinkdiss}) we find
$ \frac{\partial p_{sink}(\infty)}{\partial \Gamma}\leq 0$ and
hence that pure relaxation always yields $p_{sink}(\infty)<(N-1)^{-1}$.
Here, the losses to the environment always offset any increase in
the transport efficiency due to dephasing of the invariant states.
For the case of both pure dephasing and relaxation, the transfer
is always sub-optimal as losses to the environment always lead to
$p_{sink}(\infty)<1$.

Interestingly, it should be noted that the conclusion that pure
dissipation does not enhance transport to the sink is not always
valid. To see this, consider a system as in Fig.~(\ref{fig4}(a))
that is initially prepared in one of the invariant states such as
$(|1\rangle-|2\rangle)/\sqrt{2}$ so that one finds $p_{sink}(\infty)=0$ for
$\Gamma=0$. Now let us assume that only site 2 suffers dissipation,
$\Gamma_2>0$. The time evolution is now composed of two
contributions. In one part, the excitation is lost to the
environment leading to no successful transport. The other
part in which the exciton is not being lost is more interesting.
This branch of the time evolution is  described by a conditional
Hamiltonian that is the original Hamiltonian supplemented, crucially,
with the non-Hermitean additional term $-i\Gamma_2|2\rangle\langle 2|$
\cite{PlenioK98}.
Then the initial state $(|1\rangle-|2\rangle)/\sqrt{2}$ is not
invariant under the conditional time evolution anymore
and the relative weight of state $|2\rangle$ increases compared to
state $|1\rangle$. This in turn affects the destructive interference
for transitions to level $|3\rangle$ as the interfering amplitudes
then do not have the same weight anymore. This releases the trapped
population and leads to some transfer to site $|3\rangle$.

Thus in the
absence of pure dephasing, it is possible to find instances of
both positive and negative contributions to the transport as a
result of relaxation.
\subsection{Line broadening as a classical effect}
So far we have placed an emphasis on the effect of suppression
of destructive interference, hence addressing the wavelike feature
of the exciton transport. Dephasing may also be seen to broaden
the resonance lines of the individual sites. This effect has a
purely classical component as well. Indeed, fluctuating excitation
energies of sites leads to stronger coupling between sites whose energy
difference becomes small or even changes their energy ordering
thanks to fluctuations in time \cite{plenio08}.
\begin{figure}[t]
\centerline{\includegraphics[width=.48\textwidth]{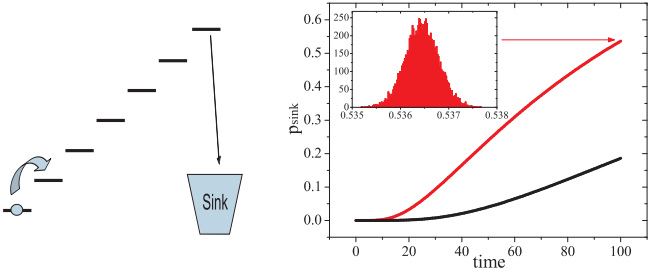}}
\caption{Transport in a classical ladder like system of $N=7$
sites, with steps of unit size ($\Delta E_{i,i+1}=1$), with a
dissipation term (with rate $\Gamma_{N+1}$) into a sink, connected
to the site $7$. The energy levels fluctuate with a Lorentzian
distribution (with width $1$) to take into account line broadening
effects and a symmetric hopping rate given by $f(j,j+1)=1/(4\pi) \
1/({\Delta E_{j,j+1}}^2 +1/16)$ between the sites $j$ and $j+1$
and $\Gamma_{N+1}=1$ (red line), while also the noiseless case is
shown (black line). Inset: Pdf distribution for $p_{sink}$ for
$t=100$ over a sample of $10^4$ instances in both hopping
scenarios. Notice that it is very narrow around the central
$p_{sink}$ value. In both cases, it is clearly shown that line
broadening dramatically enhances the transport.}\label{fig6}
\end{figure}
In order to clearly separate this mechanism from possible quantum
interference effects discussed before, we exemplify it here using
the purely classical model of $N$ sites in an ascending ladder of
sites as displayed in Fig~(\ref{fig6}) with the highest energy
site connected to a sink. The probability of finding the
excitation on the site $i$, $i=1,\ldots,N$ at a time $t$,
$p_i(t)$, is given by the Pauli master equation as \ben
  \frac{\mathrm{d}p_i(t)}{\mathrm{d}t}&=&k_{i,i+1}\,p_{i+1}(t)+
  k_{i,i-1}\,p_{i-1}(t)\\
  && -\left(k_{i+1,i}+k_{i-1,i}+\Gamma_{N+1}\delta_{iN}\right)
  p_i(t)
  \nonumber \\
  \frac{\mathrm{d}p_{sink}(t)}{\mathrm{d}t}&=& \Gamma_{N+1}\,p_N(t) \;,
\een where $k_{i,j}$ are the hopping rates from sites $j$ into
site $i$, respectively, $\Gamma_{N+1}$ is the irreversible decay
from the site $N$ into the sink and we assume that the energy
differences between neighboring sites is $\Delta E_{i,i+1}=1$. We
take account of line broadening by adding to this level spacing a
noise term leading to a Lorentzian distribution of energy levels
with width $1$. Now, we consider two very closely related
scenarios: namely (i) we consider thermal hopping, i.e.
$k_{i,j}/k_{j,i}= e^{-\Delta (E_{i}-E_{j})/ T}$ and (ii) the
hopping is symmetric $k_{i,j}=k_{j,i}$, with an hopping rate
obeying a Lorentzian in the energy difference of two neighboring
sites. Notice that the latter is perhaps closer to the quantum
case where the coupling between neighboring levels with an energy
mismatch is better approximated by case (ii). In Fig.~(\ref{fig6})
the dramatic rise in transport due to line broadening is shown for
the case of Lorentzian hopping rates. A similar behaviour was
found also for other cases. Hence, we observe that energy mismatch
in a system with static disorder inhibits the transfer of an
excitation up the ladder and the transport may be supported by
addition of dephasing.
\section{Correlated Noise}
The considerations so far have assumed Markovian local noise. In
photosynthetic complexes that will be discussed in the following
section it is however known that such a model is not fully
accurate both due to the small size of the complex compared to the
correlation length of the relevant bath and due to the fact that
the time scales for interactions may be of the order of the bath
correlation time. To gain an understanding of how spatial
correlations affect decoherence, we study a Markovian master
equation
\begin{equation}
        \frac{d\rho}{dt} = -i[H,\rho] + {\cal L}_{deph}(\rho)
        + {\cal L}_{diss}(\rho) \;,
        \label{masternonMark}
\end{equation}
where the Hamiltonian and the dissipative term are chosen as Eq.
(\ref{Hamiltonian}) and Eq. (\ref{dissipation}) respectively while
the Lindblad operator describing the correlated dephasing is given
by
\begin{eqnarray}
        \label{nolocdiss}
        {\cal L}_{deph}(\rho) &=& \sum_{mn} \gamma_{mn}\left(
        - \sigma_m^{+}\sigma_m^{-}\sigma_n^{+}\sigma_n^{-}\rho
        - \rho\sigma_n^{+}\sigma_n^{-}\sigma_m^{+}\sigma_m^{-}
        \right. \nonumber\\
        && \left. \hspace*{0.7cm}
        + \sigma_m^{+}\sigma_m^{-}\rho\sigma_n^{+}\sigma_n^{-}
        + \sigma_n^{+}\sigma_n^{-}\rho\sigma_m^{+}\sigma_m^{-}
        \right). \;\;\;\;\;\;
\end{eqnarray}
Note that the Hermitian matrix of the $\gamma_{mn}$ must be
positive semidefinite to describe a completely positive dynamics.
The effect of the noise correlations can now be seen quite clearly
by considering the special case of a two site system where
$\gamma_{11}=\gamma_{22}=\gamma$, $\gamma_{12}=\gamma_{21} =
\alpha\gamma$, $\Gamma_{k}=0$ for $k=1,\ldots,N$ and $H=0$. With
the initial state $\rho = \frac{1}{2}(|1\rangle-|2\rangle)(\langle
1|-\langle 2|)$, which led to destructive interference in
Fig.~(\ref{fig4}), we find
\begin{equation}
        \frac{d\rho}{dt} = 2(1-\alpha)\gamma (\ket{2}\bra{2}
        + \ket{2}\bra{1}+ \ket{1}\bra{2}
        + \ket{1}\bra{1}) \;.
\end{equation}
For $\alpha=1$, i.e. perfectly correlated noise, we find
${\dot\rho}=0$, i.e. the state $\rho$ remains stable. For
anti-correlated noise $\alpha=-1$ however this state would be
decaying more rapidly than under uncorrelated noise. Hence,
spatial correlations in the dephasing noise may act both to
stabilize or destabilize specific coherent superpositions and the
related destructive interference effects explained earlier in this
work. Generally, of course the spatial correlations in the noise
will not be perfect and depend for example on the distance between
sites -- more distant sites will tend to suffer weaker
noise-correlations. Needless to say, the above master equation
picture does not encompass all possible environments and also
neglects non-Markovian effects which may however be easily
included for example by using time convolved master equations
with a memory kernel that nevertheless preserve complete
positivity \cite{Lidar}. A thorough discussion of such
effects goes well beyond the scope of this work as it would
require us to take into account the details of the environment in
our simulations. However, the present paper is aimed at clarifying
fundamental mechanisms and we will therefore present a detailed
study of the environment in a future publication.
\section{Light harvesting - The FMO Complex}
\subsection{Excitation transfer}%
\begin{figure}[t]
\centerline{\includegraphics[width=.45\textwidth]{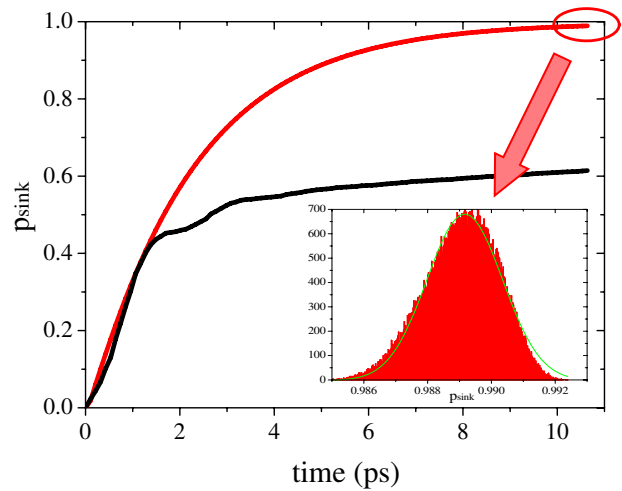}}
\caption{$p_{sink}$ vs. time (in $\mathrm{ps}$) for the FMO
complex. We show the noiseless case (black line) and the transfer
for optimal dephasing (red line). The optimal dephasing values
are given by by $\{0.157,9.432,7.797,9.432,7.797,0.922,9.433
\}~\mathrm{ps}^{-1}$. The inset shows the very narrow
probability distribution of transfer probabilities at $t\sim
10~\mathrm{ps}$ from $10^5$ samples, in the case of optimal
dephasing but with a $20\%$ static disorder in energy and coupling
rates. This suggests that dephasing assisted enhanced transport is
also robust against variations in the system
parameters.}\label{fig7}
\end{figure}

Having elucidated the fundamental mechanisms that lead to enhanced
transport in abstract, fully connected, networks, we will now
consider them in the context of transport properties in the FMO
complex~\cite{olson04} that has been the subject of several
theoretical \cite{plenio08,aspuruguzik08,Mukamel08} and
experimental \cite{fleming07a} studies recently. We will consider
both local and non-local noise and will also compare the two
settings to assess the impact of noise correlations. The FMO
complex is a pigment-protein complex that funnels the excitation
energy from the light-harvesting chlorosomes to the bacterial
reaction center in green sulfur bacteria. It is a trimer of three
identical units, each composed of seven chlorophyll $a$ molecules
embedded in a scaffolding of protein molecules. We model the FMO
complex as a completely connected network, albeit with site
dependent coupling strengths and site energies \cite{adolphs06}.
Specifically, the dynamics describing the FMO complex is composed of a
Hamiltonian part describing the coherent dynamics where the site
energies and coupling constants have been taken from tables 2 and
4 of reference \cite{adolphs06}. We then find, in matrix form
\begin{equation}
        H \!=\!\! \left(\!\!\begin{array}{rrrrrrr}
         215   & \!-104.1 & 5.1  & -4.3  &   4.7 & -15.1 &  -7.8 \\
        \!-104.1 &  220.0 & 32.6 & 7.1   &   5.4 &   8.3 &   0.8 \\
           5.1 &   32.6 &  0.0 & -46.8 &   1.0 &  -8.1 &   5.1 \\
          -4.3 &    7.1 &\!-46.8 & 125.0 &\! -70.7 &\! -14.7 &  -61.5\\
           4.7 &    5.4 &  1.0 & \!-70.7 & 450.0 &  89.7 &  -2.5 \\
         -15.1 &    8.3 & -8.1 & -14.7 &  89.7 & 330.0 &  32.7 \\
          -7.8 &    0.8 &  5.1 & -61.5 &  -2.5 &  32.7 & 280.0
          \end{array}\!\!
        \right)
        \label{hami} \nonumber
\end{equation}
where the zero of energy has been shifted by $12230$ for all sites,
corresponding to a wavelength of $\cong 800~\mathrm{nm}$ (all
numbers are given in units of cm$^{-1}=$$1.988865\cdot
10^{-23}~\mathrm{Nm} = 1.2414\,10^{-4}~\mathrm{eV}$). The
non-unitary part of the evolution is then described by
Eq.(\ref{masternonMark}). Recent work \cite{adolphs06} suggests that
it is this site $3$ that couples to the reaction centre at site 8.
For this rate, somewhat arbitrarily, we chose $\Gamma_{3,8} =
62.8/1.88~\mathrm{cm}^{-1}$, corresponding to about
$6.283~\mathrm{ps}^{-1}$ (note the equivalence $\hbar \sim 5.3$
cm$^{-1}\;\mathrm{ps}$). The measured lifetime of excitons is of the order of
$1~\mathrm{ns}$, which determines a dissipative decay rate of
$2\Gamma_k = 1/188~\mathrm{cm}^{-1}$ and that we assume to be the
same for each site \cite{adolphs06}. \\
As shown earlier, a
completely coherent dynamics is often not most ideal for the
transfer of excitons from the chlorosomes to the reaction center
and does not match with experimentally observed transfer
efficiencies. Indeed, the transfer in the FMO complex
\textit{Prosthecochloris aestuarii} undergoing completely coherent
dynamics caps at about $57\%$, shown in black in
Fig.~(\ref{fig7}). The same figure shows the effect of local
dephasing on all the sites. The enhanced transport is obtained
with dephasing rates that were optimized numerically for a
transfer time of about $5~\mathrm{ps}$~\cite{adolphs06,plenio08}.
The dynamical simulation of energy transfer in our model of the
FMO complex is shown online through two movies in \cite{movie}.

In the following we use the basic principles that we have
formulated in the previous sections to understand where and how
dephasing will be beneficial for the exciton transport in the FMO
complex. These expectations will then be corroborated by
determining numerically the optimized local dephasing rates that
will indeed be broadly in-line with the choices suggested below.
When studying the dephasing-free evolution, one observes a
behaviour quite similar to that shown in the black curve of
Fig.~(\ref{fig7}). After a rapid rise in transfer probability the
system then enters a phase of slowed transfer that is due to the
system having evolved into a state that is almost invariant in
time, i.e. it belongs to an approximate (due to the non-uniform
nature of the FMO complex) invariant subspace whose concepts have
been discussed here in the context of fully connected networks.
Indeed, a numerical simulation shows that after a time $t\sim
1~\mathrm{ps},$ the system rapidly evolves into a pure state where
the main population is approximately evenly distributed in the
sites $1$ and $2$ and where the amplitudes have a relative phase
close to $\pi$ which in turn leads to destructive interference for
transitions to level $3$. It is this destructive interference that
needs to be disrupted by a moderate level of dephasing noise on
the sites $1$ and $2$. Indeed, for $t=5~\mathrm{ps},$ the
noiseless evolution yields $p_{sink}=0.566$, while dephasing noise
only on site $1$ with $\gamma_{11}= 2.985~\mathrm{ps}^{-1}$ yields
$p_{sink}=0.730$, while dephasing noise only site $2$ with
$\gamma_{22}=20.111~\mathrm{ps}^{-1}$ yields $p_{sink}=0.772$.

Furthermore, it is crucial to note that the energy of site $3$
differs from that of sites $1$ and $2$ to an extent that the
Hamiltonian matrix elements between sites $1$ and $2$ are not
sufficient to lead to strong transfer to site $3$. This is why
a moderate amount of dephasing on site $1,2$ or $3$ may assist
the transfer process using line broadening to strengthen this
transfer channel. Indeed, for $t=5~\mathrm{ps},$ dephasing only
on site $3$ alone, of $\gamma_{33}=35.623~\mathrm{ps}^{-1}$,
already yields $p_{sink}=0.658$.

It should be noted that it is not desirable for the system to
transfer population into the sites $4,5,6$ and $7$ as this will
lengthen the time until eventual transfer into the reaction center
via site $3$. Hence strong dephasing on those sites would be ideal
to suppress this undesired exploration of the sites $4,5,6$ and
$7$. As both sites $1$ and $2$ are most strongly coupled to level
$6$, amongst these levels the strongest dephasing would be
expected for that site. For $t=5~\mathrm{ps},$ a dephasing on site
$4$ alone, with rate $\gamma_{44}=20.551~\mathrm{ps}^{-1}$, would
yield $p_{sink}=0.644$, while dephasing on site $5$ alone, with
rate $\gamma_{55}=32.076~\mathrm{ps}^{-1}$, would yield
$p_{sink}=0.718$, and dephasing on site $6$ alone, with rate
$\gamma_{66}=25.43~\mathrm{ps}^{-1}$, would yield $p_{sink}=0.626$.

We have also determined numerically the optimized dephasing rates
for $t=5~\mathrm{ps},$ where the optimal dephasing rates yield
$p_{sink}=0.903$ as compared to $p_{sink}=0.566$ in the dephasing
free case. Those are
given by $\{0.157,9.432,7.797,9.432,7.797,0.922,9.433
\}~\mathrm{ps}^{-1}$.

Finally, we would like to explore briefly to what extent non-local,
i.e. correlated, noise, may lead to further enhancements of the
transport performance of the system. To this end, we perform a
numerical optimization for $t=5~\mathrm{ps},$ now making use of
all the $\gamma_{mn}$ in Eq. (\ref{nolocdiss}) rather than simply
the diagonal elements that describe local dephasing only. We find
that with non-local dephasing noise the best transport probability
into the reaction center is given by $p_{sink}=0.931$ as compared
to $p_{sink}=0.903$ without correlations. This small additional
transfer probability suggests that, while non-local effects may
play some role for exciton transfer in the FMO complex, it is not
expected to lead to decisive improvements in this setting.
As a final remark, the basic mechanisms that we have elucidated in
previous sections all suggest a weak or moderate dependency on the
noise level where only vanishing or extremely strong dephasing
noise lead to a deterioration of performance. For a broad
intermediate level of noise and static disorder, the effects of
suppression of destructive interference and the enhanced coupling
due to line broadening are active. This is corroborated here where
we show that these optimal dephasing values are highly resilient
under variations in the system parameters, like site energies,
coupling strengths and dephasing rates. This is proved by the fact
that enhanced transport in the FMO complex remains essentially
unaffected in the presence of random variations ($20\%$) in site
and bond energies, as shown in Fig.~(\ref{fig7}). As already
demonstrated in Fig.~(\ref{fig5a}) for the fully connected
network, a numerical study of the FMO complex also shows that
variations in the dephasing rates around the optimal values do not
affect the transfer probabilities significantly. This strongly
suggests that the experimental results recorded for samples at
$77\;\mathrm{K}$ would also be observable at higher temperatures.
\subsection{Entanglement dynamics}
\begin{figure}[b]
\centerline{\includegraphics[width=.48\textwidth]{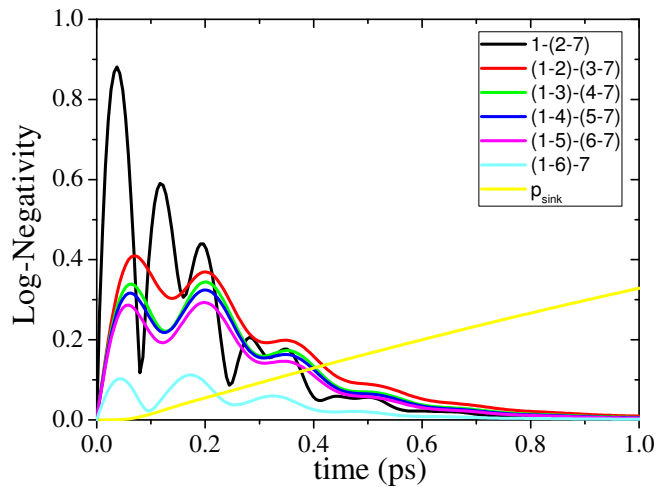}}
\caption{Entanglement, as quantified by the logarithmic negativity
\cite{enta}, across different bipartitions, as specified in the inset, for optimized
local dephasing rates at each site. The curves are for the 6 splits of the form
$(1,\cdots,k)$-vs-$(k+1,\cdots,7)$, with $k \in\{1,\cdots,6\}$, of the FMO complex sites.
Under the action of local dephasing, entanglement in our model persists up to $1\;\mathrm{ps}$ and therefore quantum correlations are dissipated well before the excitation transfer (yellow line) is
completed.}\label{enta1}
\end{figure}
Let us consider the presence of entanglement
in the FMO complex in the presence of dephasing noise.  We study the dynamics of a single excitation, initially injected in site $1$, employing
the master equation and Hamiltonian  introduced in the
previous subsections, in both the cases of local and spatially correlated dephasing.
For the case of local dephasing, we use the optimized dephasing rates as above, while for spatially correlated case the diagonal elements of the matrix $\gamma_{mn}$ are again the optimized local dephasing rates and the off-diagonal elements are random numbers.
Results for the local dephasing case are
presented in Fig.~(\ref{enta1}).
The initial exciton delocalization rapidly creates entanglement,
quantified by the logarithmic negativity of the
state, between site 1 and the rest of the complex (black line),
which degrades as time elapses. Let us remind that the logarithmic negativity of a state $\rho$ of a bipartite system $AB$ is defined as $\log||\rho^{\Gamma_A}||_1$, where $\Gamma_A$ is the partial transpose operation with respect to the subsystem $A$ and $|| \cdot ||_1$ denotes the trace norm \cite{enta}.
There is also entanglement
across other bipartitions, with a profile that is consistent
with the expectation of limited spatial distribution of the
quantum correlations. The time evolution under spatially correlated noise is broadly comparable
to the uncorrelated case shown in Fig.~(\ref{enta1}), although there is slightly more and longer lived entanglement in the correlated case. These results are in qualitative agreement with
recent studies on entanglement dynamics in bipartite qubit systems subject to forms of correlated (non-Markovian) noise \cite{nonMarkov}.

Notice, however, that the constraint on the matrix $\gamma_{mn}$ (which has to be positive semidefinite), i.e. $\gamma\ge 0$, is essential here, while violation of this constraint leads to a time evolution in which entanglement persists for much longer time but it
does not preserve positivity and is hence not physical. This highlights the
importance of considering only time evolutions that are completely positive
and trace preserving and that even minor violations may lead to significant
deviations in the entanglement dynamics.

The present analysis was carried
out in the single exciton sector in which the presence of
coherence (off-diagonal elements) is equivalent to the
presence of entanglement. A more detailed analysis, relaxing
this condition and studying the dynamics under realistic laser
excitation as well as driving from the antennae complexes
in the presence of spatial and temporal noise correlations
will be presented elsewhere \cite{usENTA}.

\section{A transparent physical picture}
The purpose of this section is to summarize the finding of previous sections and present, in a qualitative and transparent manner,
the mechanisms that may underlie the dephasing assisted transport of
excitations in quantum mechanical models of fully connected
networks and, in particular, in light-harvesting complexes.
\subsection{Destructive interference suppression}
\begin{figure}[b]
\centerline{\includegraphics[width=.46\textwidth]{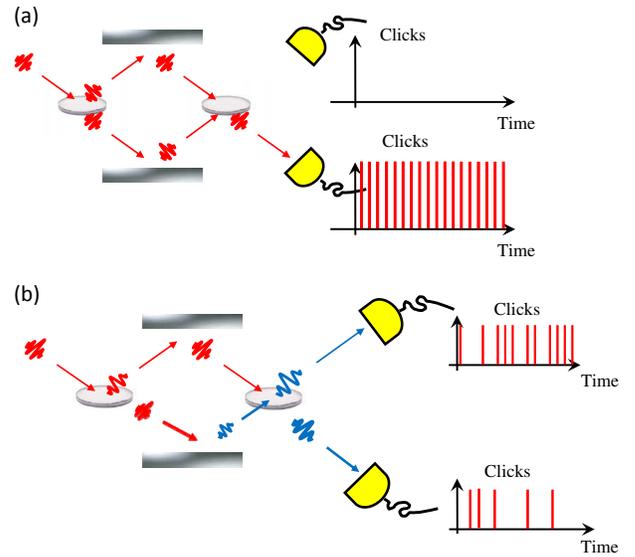}}
\caption{(a): A single photon entering a balanced Mach-Zehnder
interferometer will always emerge in one output port as the path
to the upper detector is blocked by destructive interference. (b): If the interferometer is unbalanced, as a result
of a noisy process, the condition for destructive interference is inhibited and photodetections are observed in both ports.
Thus, noise may
open additional paths for propagation. }\label{fig1}
\end{figure}
The crucial importance of destructive
interference between pathways and its suppression by dephasing
noise can be illustrated by a fundamental experiment (see
upper part of Fig.~(\ref{fig1})) that is capable of elucidating many
of the basic aspects of quantum mechanics: the balanced
Mach-Zehnder interferometer \cite{BW}. With a photon incident on
the upper input port of a 50/50 beamsplitter, one finds that it
will {\em always} emerge in the lower output port of the second
beam-splitter while it is destructive wave-interference that
prevents the photon to emerge in the upper port. This destructive
wave-interference may be perturbed in two fundamental ways, each
of which serves to extract the which-path information about the
photon~\cite{englert96,paterson} - see lower part of Fig.~(\ref{fig1}).
Firstly, local dephasing in one
of the arms of the Mach-Zehnder interferometer (e.g. random
fluctuations of the refractive index) will vary the phases and
therefore inhibit the destructive interference. Secondly, when the photons in the
beam-splitter suffer a path-dependent deterministic frequency,
time or polarization shift, then
wave-interference is inhibited because it is now possible to
identify which path was taken. The essential insight gained by
this simple example is the fact that decoherence and static
disorder, by inhibiting destructive interference, may have the
effect of opening up additional pathways for propagation in the
system (here the photon ending up either in the upper or the lower
output port).
\subsection{Line broadening}
\begin{figure}[t]
\centerline{\includegraphics[width=.48\textwidth]{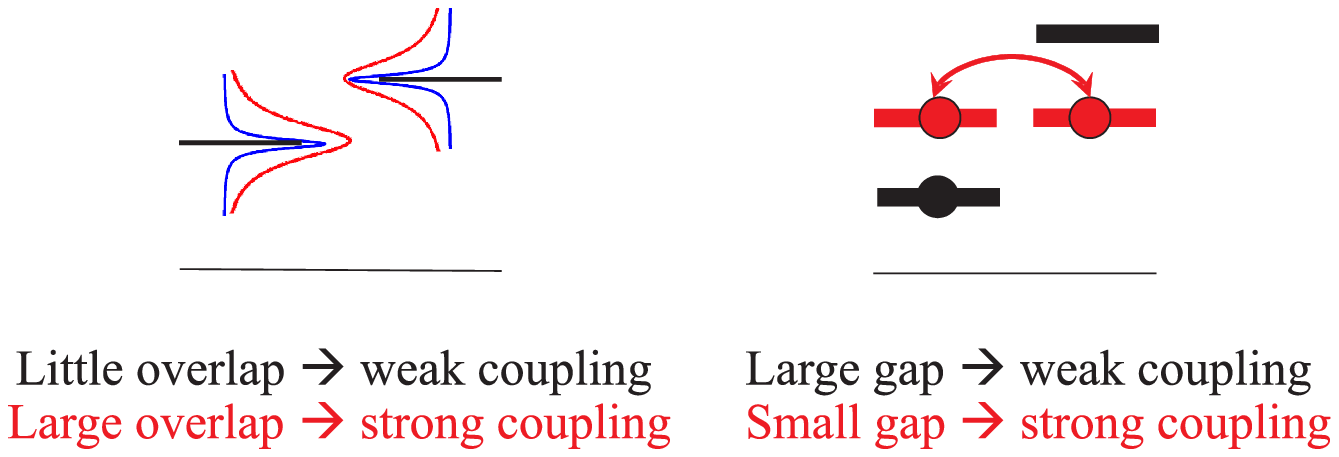}}
\caption{Left: Due to energy fluctuations, dephasing leads to a
broadening of energy levels and hence increased overlap between
sites. Right: Viewing these fluctuation dynamically, one finds
that the energy gap between levels varies in time. A non-linear
dependence of the transfer rate on the energy gap may therefore
lead to an enhancement of the average transfer rate in the
presence of dephasing noise. }\label{fig2}
\end{figure}

In addition to the inhibition of wave-like interference, dephasing
has another effect that may already be seen to be relevant on a
classical level but which can also be expected to play a role in
quantum systems. Indeed, dephasing may arise from the random
fluctuation of the on-site energies which, in turn, implies that
the energy gap between neighboring sites will fluctuate. As the
effective coupling rate between sites usually depends non-linearly
on the energy difference between neighboring levels, these
fluctuations may indeed have the effect of enhancing the average
transport rate between sites (r.h.s. of Fig.~(\ref{fig2})).
Another way of expressing this observation is by saying that dephasing leads
to line broadening so that excitation lines of interacting sites
may overlap more strongly and hence enhance transport (l.h.s. of
Fig.~(\ref{fig2})). Hence, again, propagation paths that may not
be open, or are weak, in the noise free case may be opened up
(enhanced) in the presence of dephasing.\\
\section{Concluding remarks}
We have analysed transport assisted by dephasing, and more
general types of noise, in dissipative networks such as
those describing the transport of excitons in photosynthetic
complexes \cite{plenio08,aspuruguzik08} and also considered
the presence of entanglement. We have introduced an
analytical technique that allows for the identification of
invariant subspaces of network Hamiltonians and therefore
the possibility of supporting noise assisted transport. We
have further clarified the fundamental mechanisms that lead
to this phenomenon and provided simple examples to reveal their
basic character. Of key importance are the suppression of destructive
interference by dephasing noise, static disorder in energy levels
and differences in propagation speed along different paths on the
one hand, and the enhanced overlap between sites with dephasing
broadened lines on the other. The fundamental character of these
processes suggests that they play a role in a wide variety of
transport processes whose physical and biological realisations
extend well beyond the specific example of photosynthetic complexes
discussed here. Our results suggest that Nature may actively utilize the
fundamental sensitivity of quantum systems to noise and that the
way Nature seems to exploit environmental noise is intrinsically
robust to changes in the properties of the dissipative quantum
network. This robustness corroborates the idea of the general
applicability of the described underlying principles in a variety
of transport networks and the possibility to exploit these
noise-assisted processes for achieving robust and efficient
energy transfer in artificial nano-structures.

\begin{acknowledgments}
This work was supported by the EPSRC QIP-IRC and the EPSRC grant
EP/C546237/1, the EU STREP project CORNER, the EU Integrated project
on \emph{Qubit Applications} QAP and the Royal Society via a Wolfson
Research Merit Award (MBP). We are grateful to Shashank Virmani, Graham Fleming, Tobias Brandes, Alexandra Olaya-Castro and
Al\'an Aspuru-Guzik and his group for their comments on the
manuscript.
\end{acknowledgments}

\appendix

\section{Complete solution of the master equation}

\subsection{Calculating the population transfer}
We will now present a solution of the Markovian master equation
for a fully connected network (FCN) with
$\Gamma_{i}=\Gamma,\gamma_{i}=\gamma$, and $\omega_{i}=\omega$. In
terms of the density matrix elements in the site basis
$\rho_{ij}(t)$, the equations of motion are,
\begin{eqnarray}
 \dot{\rho}_{ij}&=&-\left[2\Gamma+\Gamma_{N+1}(\delta_{iN}+\delta_{jN})+
 2\gamma -2\gamma \delta_{ij}\right]\rho_{ij} \nonumber\\
 &+& iJ\Big(\sum_{l\neq j}\rho_{il}-\sum_{l\neq i}\rho_{lj}\Big)\; ,\\
 \dot{\rho}_{00}&=& 2\sum_{j=1}^N\Gamma_j\rho_{jj}\; , \nonumber
\end{eqnarray}
where $\rho_{00}$ is the environmental population, the dots
represent time derivatives, and the time labels, e.g. $(t)$, have
been suppressed in order to neaten the presentation of this and
all subsequent equations of motion. We now introduce a set of
collective variables given by
 \be
 R_i = \sum_{j=1}^{N}\rho_{ij}\;\;\;\;\mbox{and}\;\;\;\
 \Lambda =\sum_{i=1}^N R_i = \sum_{i,j=1}^N\rho_{ij} \; ,
 \ee
and reexpress the equations of motion in terms of these collective
variables, i.e.,
\begin{eqnarray}
\dot{\rho}_{ii}&=&-2\Gamma\rho_{ii}+iJ (R_{i}-\bar{R}_{i}) \; ,\hspace{2.cm} i \neq N \nonumber\\
\dot{\rho}_{ij}&=&-2(\Gamma+\gamma)\rho_{ij}+iJ (R_{i}-\bar{R}_{j}) \; , \hspace{0.5cm} i \neq N, j\neq N \nonumber\\
\dot{\rho}_{iN}&=&-(2\Gamma+2\gamma+\Gamma_{N+1})\rho_{iN}+iJ(R_{i}-\bar{R}_{N})\; ,\\
\dot{\rho}_{NN}&=&-2(\Gamma_{N+1}+\Gamma)\rho_{NN}+iJ(R_{N}-\bar{R}_{N}) \; , \nonumber\\
\dot{\rho}_{00}&=& 2\Gamma\;\tr\rho \; . \nonumber
\end{eqnarray}
 where $\bar{R}_{i}$ denotes the complex conjugate of
 $R_{i}$. These equations are supplemented by the equations of motion for
$R_{i}$ and $R_{N}$, i.e.,
\begin{eqnarray}
\dot{R}_{i}&=&-iJ\Lambda+iJNR_{i}-2(\Gamma+\gamma)R_{i}-\Gamma_{N+1}\rho_{iN}\nonumber\\
&& + 2\gamma\rho_{ii},
\end{eqnarray}
\begin{eqnarray}
\dot{R}_{N}&=&-iJ\Lambda+iJNR_{N}-(2\Gamma+2\gamma+\Gamma_{N+1})R_{N}\nonumber\\
&& +(2\gamma-\Gamma_{N+1})\rho_{NN} \; ,
\end{eqnarray}
and the equation of motion for $\Lambda$, i.e.,
\begin{equation}
\label{sigmadot}
\dot{\Lambda}=-2(\Gamma+\gamma)\Lambda-\Gamma_{N+1}(R_{N}+\bar{R}_{N})+2\gamma\;\tr\rho
\;.
\end{equation}
The final term is the total population in the network, which is
not preserved due to the losses to the environment and sink.
However the following relation must always hold, i.e.
 \be
 \label{prob}
1= \tr\rho + \rho_{00} + p_{sink} \; .
 \ee
Note now that the collective variables $R_{i}$ will in general be
complex, while $\Lambda$ is always real. In particular, let \be
 R_N = X + i Y \; ,
 \ee
and this results in the following differential equations which
form a closed system:
 \ben
\dot{\Lambda}&=&-2(\Gamma+\gamma)\Lambda-2\Gamma_{N+1}X+2\gamma(1-\rho_{00}-p_{sink}) , \nonumber\\
\dot{X}&=&-(2\Gamma+2\gamma+\Gamma_{N+1})X+ (2\gamma-\Gamma_{N+1})\rho_{NN} -JNY , \nonumber\\
\dot{Y}&=&-(2\Gamma+2\gamma+\Gamma_{N+1})Y + JNX -J\Lambda , \nonumber\\
\dot{\rho}_{NN}&=&-2(\Gamma+\Gamma_{N+1})\rho_{NN}  -2JY , \nonumber\\
\dot{\rho}_{00}&=&2\Gamma(1-\rho_{00}-p_{sink}) , \nonumber\\
\dot{p}_{sink}&=&2\Gamma_{N+1}\rho_{NN} \;. \nonumber
 \een
The initial conditions considered in the main text are
 \be
\Lambda=1,\;X=0,\;Y=0,\;\rho_{NN}=0,\;\rho_{00}=0,\;p_{sink}=0\;
.\nonumber
 \ee
This system of coupled differential equations can be converted
into a set of algebraic equations via the Laplace transform
$\mathcal{L}$. Decoupling these equations, one finds that the
Laplace transforms of the dynamical variables contain only simple
poles and the inverse Laplace transformation can be performed
analytically. The analytical results presented in the paper are
all derived from this procedure. In particular, the full dynamical
problem can be reduced to the solution of the following set of
equations for the Laplace $s$-domain variables
$\tilde{\Lambda}=\mathcal{L}[\Lambda(t)]$ , i.e.,
 \begin{footnotesize}
 \ben
  \label{lapeqns}
 (s+2\Gamma+2\gamma)\tilde{\Lambda} + 2\Gamma_{N+1}\tilde{X}+2\gamma\tilde{p}_{sink}+2\gamma\tilde{\rho}_{00}-2\gamma/s-1 = 0 \; , \nonumber\\
 (s+2\Gamma+2\gamma+\Gamma_{N+1})\tilde{X} + (\Gamma_{N+1}-2\gamma)\tilde{\rho}_{NN}+ JN\tilde{Y} = 0 \; , \nonumber\\
 (s+2\Gamma+2\gamma+\Gamma_{N+1})\tilde{Y} + J\tilde{\Lambda}- JN\tilde{X} = 0 \; , \nonumber \\
 (s+2\Gamma+2\Gamma_{N+1})\tilde{\rho}_{NN} + 2J\tilde{Y} = 0 \; , \nonumber \\
 (s+2\Gamma)\tilde{\rho}_{00} + 2\Gamma\tilde{p}_{sink} -2\Gamma/s = 0 \; , \nonumber\\
 s\tilde{p}_{sink} -2\Gamma_{N+1}\tilde{\rho}_{NN} = 0 \; .
 \;\;\;\;
 \een
 \end{footnotesize}
The quantity of eventual interest is $\tilde{p}_{sink}$, but the
last equation can be used to eliminate $\tilde{p}_{sink}$ in
favour of $\tilde{\rho}_{NN}$, leaving five linear equations
involving an equal number of variables. When determining the poles
of the Laplace transform of $\rho_{NN}$, one finds that the
characteristic equation is therefore one of fifth-order, and
consequently its roots cannot be generally expressed in analytical
form. However, some special examples in which the dynamics of the
full system can be analytically derived will be shown in the
following.
\subsection{The full density matrix dynamics without losses or
pure dephasing}
As an example of how the full dynamics and final state of the
system can be extracted from the system of equations given above,
we present an analysis of the case $\Gamma=0,\gamma=0$. The
coupled equations for the collective variables in the Laplace
domain given in Eqs. (\ref{lapeqns}) are reduced to
 \ben
  s\tilde{\Lambda} + 2\Gamma_{N+1}\tilde{X}-1 &=& 0 \; , \nonumber\\
 (s+\Gamma_{N+1})\tilde{X} + \Gamma_{N+1}\tilde{\rho}_{NN}+ JN\tilde{Y} &=& 0 \; , \nonumber\\
 (s+\Gamma_{N+1})\tilde{Y} + J\tilde{\Lambda}- JN\tilde{X} &=& 0 \; ,\\
 (s+2\Gamma_{N+1})\tilde{\rho}_{NN} + 2J\tilde{Y} &=& 0 \; , \nonumber\\
 s\tilde{p}_{sink} -2\Gamma_{N+1}\tilde{\rho}_{NN} &=& 0 \; . \nonumber
 \een
Note that in the absence of dissipation to the sink, the system
could be solved completely in terms of the collective variables
via just the first three of the governing equations. The
dissipation to the sink is a local process and requires that we
include $\rho_{NN}$ in the system of equations in order to find
the full dynamics. Let us point out that introducing other local
variables into the density matrix equations, e.g. pure dephasing,
local dissipation, different on-site energies etc., would
similarly require extra local variables to enter into the
collective equations of motion, e.g. the sink population and the
environmental one. For the case considered here, one finds that
the Laplace transform of $\rho_{NN}$, is given by
\begin{equation}
\tilde{\rho}_{NN}=\frac{2J^{2}(s+\Gamma_{N+1})}{(s-s_{1}^{+})(s-s_{1}^{-})(s-s_{2}^{+})(s-s_{2}^{-})}
\; ,
\end{equation}
where the $s_{i}^{\pm}$ are the four roots of the equation
\begin{eqnarray}
0&=&s(s+\Gamma_{N+1})^{2}(s+2\Gamma_{N+1})\nonumber\\
&+&J^{2}(Ns+2\Gamma_{N+1})(Ns-2\Gamma_{N+1}+2N\Gamma_{N+1})\;.\nonumber
\end{eqnarray}
The laplace transform has four simple poles, and the inversion is
now straightforward. The solution in the time domain is given in
terms of the residues of $\tilde{\rho}_{NN}$ as
\begin{equation}
\rho_{NN}(t)=\sum_{i=1,2}\left(\mathrm{Res}(\tilde{\rho_{NN}},s_{i}^{+})
\,e^{s_{i}^{+}t}+\mathrm{Res}(\tilde{\rho}_{NN},s_{i}^{-})\,e^{s_{i}^{-}t}\right)
\;.\label{rhot}
\end{equation}
The roots above can be determined analytically, and can be written
in the form,
\begin{eqnarray}
s_{1}^{\pm}=-\Gamma_{N+1}\pm\Delta_{+} \; , \nonumber\\
s_{2}^{\pm}=-\Gamma_{N+1}\pm\Delta_{-} \; , \label{Delta}
\end{eqnarray}
where
\begin{eqnarray}
\Delta_{\pm}^{2}&=&\frac{1}{2}(\Gamma_{N+1}^{2}-J^{2}N^{2}) \; ,\nonumber\\
&\pm&\frac{1}{2}\sqrt{(\Gamma_{N+1}^{2}+J^{2}N^{2})^{2}-16(N-1)\Gamma_{N+1}^{2}J^{2}}
\; .\nonumber
\end{eqnarray}

Note that $\rho_{NN}(t)$ vanishes as $t\rightarrow \infty$, as one
would expect. This is ensured by the relation
$\Gamma_{N+1}>\Delta_{\pm}$, which using Eqs. (\ref{Delta}) can be
shown to be always valid. Calculating the residues of these poles,
one finally obtains $\rho_{NN}(t)$ from Eq. (\ref{rhot}),
\begin{equation}
\rho_{NN}(t)=\frac{2J^{2}e^{-\Gamma_{N+1}\,t}\left[\cosh(\Delta_{+}t)-
\cosh(\Delta_{-}t)\right]}{\sqrt{(\Gamma_{N+1}^{2}+J^{2}N^{2})^{2}-
16(N-1)\Gamma_{N+1}^{2}J^{2}}} \; .
\end{equation}
Once $\rho_{NN}(t)$ is determined, $p_{sink}$ is obtained easily
via,
\begin{eqnarray}
p_{sink}(t)&=&2\Gamma_{N+1}\int_{0}^{t}\rho_{NN}(t')dt' \; .
\end{eqnarray}
Taking the limit $t\rightarrow \infty$,  one recovers the result
found by both numerical simulations and the invariant subspace
approach (see below), which is
 \be
p_{sink}(\infty)=\frac{1}{N-1}\nonumber \; .
 \ee

Having determined $\rho_{NN}(t)$ we can now, in theory, determine
the rest of the density matrix elements. Formally one has to
substitute the solution for $\rho_{NN}(t)$ back into the system of
equations in order to determine the collective variables
$\Lambda(t)$ and $R_{N}(t)$. Doing this explicitly is rather
tedious, and we will focus instead on some general features of
the solutions. First, consider the equation of motions for
$R_{i}(t)$ and $\rho_{iN}(t)$,
\begin{eqnarray}
\dot{R}_{i}&=&iJNR_{i}-\Gamma_{N+1}\rho_{iN}-iJ\Lambda \; , \nonumber\\
\dot{\rho}_{iN}&=&-\Gamma_{N+1}\rho_{iN}+iJ(R_{i}-\bar{R}_{N}) \;.
\nonumber
\end{eqnarray}
These can be combined to obtain a closed second-order
differential equation for $\rho_{iN}$,
\begin{eqnarray}
\ddot{\rho}_{iN}+(\Gamma_{N+1}-iJN)\dot{\rho}_{iN}-iJ\Gamma_{N+1}(N-1)\rho_{iN} \nonumber\\
=J^{2}(\Lambda-NR_{N})-iJ\dot{R}_{N}\;. \label{rhoin}
\end{eqnarray}
The initial conditions are
 \be
\rho_{iN}(0)=0,\hspace{0.5cm}\dot{\rho}_{iN}(0)=0,\hspace{0.5cm}
\rho_{1N}(0)=0,\hspace{0.5cm}\dot{\rho}_{1N}(0)=iJ \;. \nonumber
 \ee
Looking at Eq. (\ref{rhoin}), one can see that the r.h.s. is a
fixed function of the previously determined collective variables
$R_{N}$ and $\Lambda$. Given that for $i\neq 1$ all the initial
conditions are identical, all $\rho_{iN}$ must be identical
functions of time. However, the different initial conditions for
$\rho_{i1}$ mean that this coherence can be different from the
others. A similar second-order equation for $R_{i}$ can also be
obtained, showing that all $R_{i}$ are the same except for $R_{1}$
and $R_{N}$. The equations of motion for the remaining density
matrix elements are,
\begin{equation}
\dot{\rho}_{ij}=iJ (R_{i}-\bar{R}_{j}), \hspace{2cm} [i,j \neq N]
\; ,
\end{equation}
which implies that all matrix elements are identical functions of
time for $i,j\neq N$, except $\rho_{11}$ and $\rho_{1j}$. A
further implication is that $\rho_{1j}=\rho_{1k}$ for all
$j,k\neq1,N$. With these results one can obtain several results
concerning the stead-state populations and coherences. The steady
states of the collective variable equations of motion lead to the
final values,
 \be
 X=0,\hspace{0.8cm} Y=0,\hspace{0.8cm} \rho_{NN}=0,\hspace{0.8cm} \Lambda=0 \; .
 \ee
Substituting these results into Eqs. (\ref{rhoin}), one finds that
all $\rho_{iN}=0$ in the steady-state. The steady-state
populations must also obey the conservation of population sum
rule,
 \be
 1=\sum_{j=1}^{N}\rho_{jj}+p_{sink} \; ,
 \ee
which using the results $\rho_{NN}=0$, $p_{sink}=(N-1)^{-1}$, and
$\rho_{ii}=\rho_{ll}$ for $i,l\neq1,N$ implies the condition,
 \be
 \frac{N-2}{N-1}=(N-2)\rho_{ii}+\rho_{11} \; .
 \ee
Setting the time derivative $\dot{\rho}_{i1}=0,$ the following
steady-state condition must hold,
\begin{eqnarray}
(N-1)\rho_{ii}&=&\rho_{11}+(N-1)\rho_{i1} \; .
\end{eqnarray}
Setting $\dot{R}_{i}=0$ one also finds that $R_{i}=0$ in the
steady-state. This implies,
\begin{equation}
\rho_{ij}=-\frac{\rho_{i1}}{N-2}.
\end{equation}
Combining these relationships, the final populations and
coherences are
 \ben
\rho_{11}&=&\left(\frac{N-2}{N-1}\right)^{2},\;\;\rho_{i1}=
-\frac{N-2}{(N-1)^{2}}, \nonumber \\
\rho_{ij}&=&\rho_{ii}=\frac{1}{(N-1)^{2}} \; . \nonumber
 \een
\subsection{The full density matrix dynamics with losses and pure
dephasing}
In the most general case the equations of motion for the
individual matrix elements of the density matrix can be written as
\begin{eqnarray}
 \label{dmij}
 \dot{\rho}_{ij}&=&-\big[\Gamma_i+\Gamma_j+\Gamma_{N+1}(\delta_{iN}+\delta_{jN})+
 \gamma_i+\gamma_j-2\gamma_i\delta_{ij} \nonumber\\
 &+& i(\omega_i-\omega_j)\big]\rho_{ij}+
 iJ\Big(\sum_{l\neq j}\rho_{il}-\sum_{l\neq i}\rho_{lj}\Big) \; ,\\
 \dot{\rho}_{00}&=& 2\sum_{j=1}^N\Gamma_j\rho_{jj} \; .
\end{eqnarray}
Under the assumption that all the sites have equal energies,
dephasing and dissipation, Eqs. (\ref{lapeqns}) can be solved
completely in terms of a few collective variables but analytically
only in the Laplace domain. Indeed, the sink population is then
expressed, in terms of the Laplace variable $s$, as
 \be
\tilde{p}_{sink}(s)=4J^{2}\Gamma_{N+1}\frac{(s+\Gamma_a)(s+\Gamma_{b})}{s\Delta(s)}
\; ,
 \ee with
 \begin{eqnarray}
 \Delta(s)&=&(s+2\Gamma)(s+\Gamma_a)(s+\Gamma_b)^{2}(s+\Gamma_c)\nonumber\\
 &+& 4J^{2}\Gamma_{N+1}(4\gamma(s +\gamma+2\Gamma)-J^{2}\Gamma_{N+1}(s-2\gamma+2\Gamma)\nonumber\\
 &+& 4J^{2}N(s+2\Gamma) \left(\gamma(s+\Gamma_a)-\Gamma_{N+1}\gamma+\Gamma_{N+1}^2\right)\nonumber\\
 &+&J^{2}N^2(s+2\Gamma)(s+\Gamma_a)(s+\Gamma_c) \; ,
 \end{eqnarray}
 where
 \be
\Gamma_{a}=2\gamma+2\Gamma,\;\;\Gamma_{b}=2\gamma+2\Gamma+\Gamma_{N+1}
,\;\;\Gamma_{c}=2\Gamma+2\Gamma_{N+1} \; .
 \ee
We can now use the above equations to solve for the sink
population at any time, but that involves solving for the roots of
$\Delta(s)=0,$ which being a fifth order polynomial cannot be
solved in closed form. Here, we stress on the asymptotic value of
the sink population which is easily obtained via the final value
theorem and is given by
  \be
p_{sink}(t=\infty) = \lim_{s\rightarrow 0}[ s\;\tilde{p}_{sink}]=
4J^{2}\Gamma_{N+1}\frac{\Gamma_a\Gamma_{b}}{\Delta(0)} \; .
  \ee
This expression is considerably general, and in particular one can
recovers the noiseless case, i.e.
 \be
p_{sink}(\infty) = \lim_{{\begin{array}{c}
  \gamma\rightarrow 0, \\
  \Gamma\rightarrow 0 \\
\end{array}}}\left[4J^{2}\Gamma_{N+1}\frac{\Gamma_a\Gamma_{b}}{\Delta(0)}\right]=\frac{1}{N-1} \; .
 \ee
For the case of purely local dephasing, i.e. $\Gamma=0$,
one finds
 \be
 p_{sink}(\infty) =
 \lim_{\Gamma\rightarrow 0}4J^{2}\Gamma_N\frac{\Gamma_a\Gamma_{b}}{\Delta(0)}=1 \; .
 \ee
\subsection{An invariant subspace approach for FCNs}
We have shown that the full density matrix equations can be solved
exactly in a number of cases, and have used these to obtain -
somewhat laboriously - several of the results found numerically.
This section presents a more detailed development of the invariant
subspace method, which, as discussed in the main text, is a powerful
method that allows one to obtain the results given above in an
intuitive way and which can be generalised to deal with situations
that cannot be easily tackled by solving the dynamical equations.

For the case of $\Gamma_{j}=0,\gamma_{j}=0$, the initial state of
the system can be expanded in terms of invariant and non-invariant
defined in the main text. This expansion is explicitly given by
\be
 \label{expansion}
\ket{1}=\left(\frac{1}{N-1}\right)\left[\overbrace{\left(
\sum_{j=2}^{N-1}(|\psi_{j}\rangle\right)}^{\mathrm{Invariant}}+
\overbrace{\sqrt{N}|\phi\rangle -|N\rangle}^{|+\rangle}\right] \;.
 \ee
The invariant states are eigenstates of the Hamiltonian and are
decoupled from the dissipative dynamics that acts at site $N$. The
only non-trivial time development comes from the term in the
expansion that lies outside of the invariant subspace. Defining
the state $|+\rangle$ as,
$$|+\rangle=\sqrt{N}|\phi\rangle-|N\rangle,$$
one can show that the amplitudes of a state of the form
$|\psi\rangle(t)=a(t)|+\rangle+b(t)|N\rangle$ obey the following
closed system of equations of motion,
 \be i\left(
   \begin{array}{c}
     \dot{a}(t) \\
     \dot{b}(t) \\
   \end{array}
 \right)
=\left(
                                 \begin{array}{cc}
                                   J(N-2) & J \\
                                   J(N-1) & 0 \\
                                 \end{array}
                               \right)\left(
   \begin{array}{c}
     a(t) \\
     b(t) \\
   \end{array}
 \right) \; .
 \ee
These equations show that spectral weight oscillates between
$|+\rangle$ and $|N\rangle$, and will thus be entirely lost to the
sink as $t\rightarrow\infty$ due to the irreversible relaxation
which acts on site $N$. As all the weight held in the
non-invariant subspace is initially held in $|+\rangle$, we can
immediately predict from the initial state expansion that,
 \be
p_{sink}(\infty)=\frac{\langle+|+\rangle}{(N-1)^{2}}=
\frac{N-1}{(N-1)^{2}}=\frac{1}{N-1} \; .
 \ee

The arguments given above can also be adapted to analyze the
effects of local perturbations on the sites. These perturbations
may be changes to the site energies or changes in the hopping rate
between the sites and a given site. The change to the hopping rate
of site $\beta$ can be described by the new Hamiltonian,
\begin{eqnarray}
H&=&\sum_{i\neq j}J_{ij}|i\rangle\langle j| \; ,\\
J_{ij}&=&J+ \Delta J_{i\beta}\delta_{i\beta}+\Delta
J_{j\beta}\delta_{j\beta} \; .
\end{eqnarray}
We now introduce the reduced invariant subspace of the perturbed
Hamiltonian as discussed in the main text. For a network with $D$
perturbed sites described by states $|d\rangle$, the initial state
can be expanded in the invariant states $|\psi_{j}\rangle$ as,
 \be
\label{expansion}
\ket{1}=\left(\frac{1}{N-D-1}\right)\left[\tilde{\sum}_{j}|\psi_{j}\rangle+\overbrace{\sqrt{N}|\phi\rangle
-|N\rangle-\sum_{D}|d\rangle}^{|D\rangle}\right] \; .
 \ee
where the tilde on the sum denotes that all $|\psi_{j}\rangle$
which contain weight on a perturbed site are excluded. Once again
one can show that the Hamiltonian dynamics leads to oscillations
of the spectral weight initially held in non-invariant part of the
expansion of $|1\rangle$ amongst the states $|D\rangle$,
$|d\rangle$, and$|N\rangle$. The leakage to the sink through site
$N$ therefore causes the eventual transfer of all of this weight
into the sink. The final weight transferred to the sink is
therefore $p_{sink}(\infty)=(N-D-1)^{-2}\langle D|D\rangle$, which
gives
\begin{equation}
p_{sink}(\infty)=\frac{1}{N-D-1} \; .
\end{equation}
The approach outlined above, which essentially consists of writing
the initial condition in terms of degenerate, invariant (to
dissipation) states and a remaining non-stationary part, fails if
the perturbed site is the injection site. The reason is that in
this case one can no longer write the initial condition as we did
before. If one tries to expand the initial state in the invariant
states, one finds that
\begin{equation}
|1\rangle=\left[\sqrt{N}|\phi\rangle-\sum_{j\neq\alpha}|j\rangle\right]
\; .
\end{equation}
Comparing with Eq. (\ref{expansion}), one can see that there is no
invariant subspace, the wave function only contains a
non-stationary component which has precisely the form we have
argued always decays to zero. Therefore one finds that all the
excitation is transferred to the sink and $p_{sink}=1$ for any
local perturbation on the injection site $1$.
\subsection{Invariant subspaces for general networks}
Consider a network of $N$ sites with a general Hamiltonian that
has $D$ degenerate orthonormal eigenstates $|\Psi_{k}\rangle$ and
$N-D$ non-degenerate orthonormal eigenstates $|\Phi_{k}\rangle$.
We shall now present a method to construct a generalised invariant
subspace that is unaffected by dissipation at site $N$ (we assume
no other sources of noise or losses in the system). Within the
degenerate and non-degenerate manifold of states, isolate all
states with no overlap with $|N\rangle$. These states are by
definition in the invariant subspace. We now take linear
combinations of the remaining degenerate eigenstates to create a
new basis using the transformation
\begin{eqnarray}
|\tilde{\Psi}_{l}\rangle &=& \sum_{k=1}^{D'}u_{lk}|\Psi_{k}\rangle \; , \nonumber\\
u_{lk} &=& \frac{1}{\sqrt{D'}}\frac{e^{\frac{2\pi i l k}{D'}}}
{\langle N|\Psi_{k}\rangle} \; , \label{trans}\\
u_{lk}^{-1} &=& \frac{1}{\sqrt{D'}}e^{\frac{-2\pi i l
k}{D'}}\langle N|\Psi_{k}\rangle \nonumber \; ,
\end{eqnarray}
where $k$ runs over all $D'$ degenerate eigenvectors with finite
$\langle N|\psi_{k}\rangle$. The overlap of the new basis with
$|N\rangle$ is given by,
\begin{eqnarray}
\langle
N|\tilde{\Psi}_{l}\rangle&=&\frac{1}{\sqrt{D'}}\sum_{k=1}^{D'}e^{\frac{2\pi
i  l k}{D'}} \; .
\end{eqnarray}
There are $D'$ states $|\tilde{\Psi}_{l}\rangle$ where the index
$l=1 \ldots D'$. For $l<D'$ the overlap is zero as
$\sum_{k=1}^{D'}e^{\frac{2\pi i l k}{D'}}=0$, and therefore this
transformation creates $D'-1$ states in the generalised invariant
subspace with $\langle N|\tilde{\psi}_{l}\rangle=0$. The
transformation also generates one state that is outside the
subspace given by $|\tilde{\Psi}_{D'}
\rangle=D'^{-\frac{1}{2}}\sum_{k=1}^{D'}\langle
N|\Psi_{k}\rangle^{-1} |\Psi_{k}\rangle$. Thus an invariant
subspace can be generated for any network with degeneracy. This
procedure systematically generates the invariant subspace,
although it should be noted that basis is not unique as various
other linear combinations of the degenerate eigenstates can be
made. This is clear if one follows this procedure for the FCN, as
the states generated here are not the pair eigenstates but rather
delocalised states. We have also ignored the case of multiple
degenerate manifolds, although the generalisation of the procedure
to this case simply involves performing transformations of the
form given in Eq. (\ref{trans}) to each manifold.

In order to study the transport dynamics of the network, we
express the initial condition of the system $|1\rangle$ by an
expansion in the original orthonormal eigenstates of the system.
We then transform the degenerate eigenstates to the new eigenbasis
using the inverse transformation of Eq. (\ref{trans}) and group
the states into the generalised invariant and non-invariant
groups. The part of the initial condition that can be represented
by elements of the invariant subspace is unaffected by the
dissipation at state $|N\rangle$ and the initial spectral weight
will be preserved in the network. Using the transformation above,
the amount of the initial excitation that is transferred to the
bath is given by the weight held outside of the generalised
invariant subspace. The population transferred to the sink is
therefore,
\begin{eqnarray}
p_{sink}=D^{-2}\left(\sum_{k=1}^{D}\langle
N|\Psi_{k}\rangle|^{-2}\right)|\sum_{k=1}^{D}\langle
N|\Psi_{k}\rangle\langle\Psi_{k}|1\rangle|^{2} \nonumber \\
+\sum_{k=1}^{N-D}|\langle\Phi_{k}|1\rangle|^{2} \;, \nonumber
\end{eqnarray}
where it is assumed that there are no naturally occurring
invariant eigenstate, i.e. $D'=D$.


\begin{thebibliography}{99}
\bibitem{note1}
Note that in a strongly interacting system such as the FMO complex
a master equation with local Lindblad terms is not actually the
result of an interaction with local baths. In fact, the derivation
of master equations in interacting systems is fraught with
difficulties. This issue will be discussed elsewhere. A detailed
study of this and the impact of cooperative and non-Markovian
noise will be presented in a forthcoming publication.
\bibitem{note2}
A more complete expression for $p_{sink}$ in a general network is
presented in the Appendix.
\end{thebibliography}
\end{document}